# New measurement method for weak magnetic fields using magnetically induced deformation of chemical bonds in Co-CsPbBr$_3$ quantum dots

Yanyan Zhang[1], Yuan Zhang[2], Yong Pan[*2]

[1] School of Pharmacy, Hanzhong Vocational and Technical College, Hanzhong, 723000, China

[2] College of Science, Xi'an University of Architecture and Technology, Xi'an, 710055, China

* Correspondence: panyong@xauat.edu.cn

## Abstract

The research on weak magnetic field detection is of great significance in advancing the development of bioscience, aerospace, chip manufacturing and other fields. However, the weak magnetic detecting still face some problems, including the large size of the detectors and the limited detection scale. To contribute to the detection of weak magnetic fields, the Co-CsPbBr$_3$ colloidal quantum dots (QDs) composite magnetic material was synthesised on the basis of the theory of room temperature ferromagnetism, molecular polarisation and vibration level of chemical bond. The synthesis involved the mixing of Co$^{2+}$ into CsPbBr$_3$, an all-inorganic perovskite with activated ions. Subsequently, a weak magnetic field measurement system was devised, comprising working medium samples and a vibration level detection optical path. Following the acquisition, comparison, processing and analysis of multiple data sets, a Stokes displacement function model was established under different magnetic field sizes and the weak magnetic field intensity range of Pitsla (pT) was measured. The Pitsla weak magnetic field measurement system proposed in this paper provides a reference for the development of non-contact weak magnetic measurement methods and for the advancement of intelligent and low-dimensional weak signal measurement applications.

## 1. Introduction

Weak magnetic fields are defined to be below 200 Gauss, so the target measurement range of the device should be below 200 Gauss (0.02 T) [1]. A weak magnetic sensor is a device that converts the magnetic performance changes of sensitive components caused by external factors, including magnetic fields, currents, stress and strain, temperature, and light, into electrical signals [2]. These signals are then used to detect the corresponding physical quantities. Weak magnetic detectors have a wide range of applications in several fields of study. In the field of scientific research, weak magnetic detectors are employed to investigate natural phenomena, including geomagnetic and biomagnetic. In the field of medical diagnosis, weak magnetic detectors can be employed to detect magnetic field changes in organisms, including the measurement of the heart's magnetic and the detection of magnetic fields generated by the brain [3]. These techniques provide an important means for the early diagnosis and treatment of diseases. In the process of nucleic acid extraction, the magnetic bead method is frequently employed, with ultrasound assistance being used to enhance the efficiency and automation of weak magnetic detection [4]. In the field of national defence and security, weak magnetic detectors can be employed to detect targets such as mines and submarines, thereby enhancing the country's security and defence capabilities [5]. Furthermore, weak magnetic detectors can also be utilised in materials science, energy detection [6]. Different weak magnetic detectors use different magnetic sensing components and signal processing methods to meet the needs of different application scenarios. In recent years, significant progress has been made in the technology of weak magnetic detectors [7]. The research and development of new magnetic sensitive components, such as superconducting quantum interferometer (SQUID) [8], giant magnetoresistance (GMR) [9], etc., has greatly improved the sensitivity of the weak detector. Meanwhile, advances in weak fabrication technology have enabled detectors to achieve higher levels of integration and smaller sizes. In addition, the weak magnetic detector also combines other advanced technologies, such as weak electronics, optoelectronics, to form a multi-functional, high-performance detection system. However, although weak magnetic detectors have achieved remarkable results in many fields, they still face some problems including: the large size of the weak magnetic field detectors, which are mostly conventional circuit-based detectors with poor applicability; the limited detection range of weak magnetic fields, with few reports of accurate detection below the pT level.

The current trend in the field of weak magnetic field measurement is the development of miniature sensors that are small size, low power consumption, and low detection level. It is not possible to guarantee increased detection levels with reduced dimensions in the case of inherently electronic detectors. Furthermore, they lack the flexibility required for use in many special applications. Thus, new weak magnetic field measurements are urgently needed. Optical test methods are primarily employed for the assessment of radiometric and optical metrics, the evaluation of non-optical physical quantities, and the examination of optoelectronic devices, materials, and systems [10]. The introduction of the optical measurement method has reinvigorated the field of precision measurement, which is the application of optical technology in conjunction with modern electronic technology. The method offers a number of notable advantages, including non-contact operation, high automation, high sensitivity, rapid speed, high efficiency, three-dimensionality, rapidity and real-time performance [11]. All inorganic perovskite ($CsPbX_3$) is very suitable for optical detection due to their suitable forbidden band width, wide spectral absorption range and high carrier diffusion length [12]. The creation of magnetic perovskites by the introduction of magnetic elements could facilitate the development of new techniques for the measurement of weak magnetic fields. The chemical element cobalt (Co) is situated within Group VIII, Period IV of the periodic table of the chemical elements. It is frequently utilised in the production of Alnico permanent magnet alloys. In contrast to Fe and Ni, Co exhibits superior ferromagnetism, with a Curie point as high as 1150°C and a coercive force that is 2.5 times higher than that of typical magnetic materials. Additionally, it displays minimal loss of magnetism in vibrational modes [13]. The origin of $Co^{2+}$ ferromagnetism at room temperature can be attributed to the properties of its electronic structure and magnetic moments. The electronic structure of $Co^{2+}$ is formed by the arrangement of 25 electrons in the following configuration: $1s^2 2s^2 2p^6 3s^2 3p^6 3d7$. This arrangement in the generation of unpaired electrons around the nucleus, which can generate magnetic moments, thus making $Co^{2+}$ magnetic [14]. It can be reasonably deduced that the combination of room temperature ferromagnetic elements and perovskite will provide a valuable experimental reference for the detection of weak magnetic fields.

In this study, we aim to synthesis $Co-CsPbBr_3$ quantum dots and utilize the data pertaining to their structural alterations subsequent to exposure to a magnetic field for the purpose of conducting weak magnetic field measurements. Here, $CsPbBr_3$ quantum dots will be synthesize via them method of heat injection, $CoBr_2$ will be added in the process to prepare $Co-CsPbBr_3$ quantum dots, and then the mixed material will be placed in a magnetic field. The Raman spectrum will be used to get information about corresponding bond vibration at different magnetic field environment, which could indirectly prove the effectiveness of chemical change incorporation. Then, the magnetic field will be calibrated based on the information obtained from the Raman spectra of the material to achieve magnetic field detection.

## 2. Experimental

**Chemicals.** Cesium carbonate ($CaCo_3$, 99.9%, purity), octadecene (ODE, 90%, purity), oleic acid (OLA, 90%, purity), oleamine (OAm, 50%, purity), lead bromide ($PbBr_2$, 99%, purity), cobalt bromide (99%, purity), n-hexane (97%, purity).

**Synthesis of $Co-CsPbBr_3$ QDs.** Add 33.36 mg of lead bromide 27.34 mg of cobalt bromide, and 20 ml of octadecene to the dried 50 ml three-neck flask, heat to 100-150 °C and stir, after heating for 45 minutes, add 7.5 ml of oleic acid and 7.5 ml of oleamine. After 30 minutes of continuous heating, then rise to 150-200 °C, at which point 0.4 ml of precursor solution is quickly injected into the three-neck flask with a pipette, and after 10 seconds, the three-neck flask is placed in an ice water bath to quickly cool to room temperature. A schematic diagram of the process is shown in Figure 1a. At this point, the resulting crude solution is centrifuged and purified at 3000 rpm for 30 minutes, after centrifugation, the supernatant is discarded, and the pellet is dispersed in n-hexane. The resulting n-hexane-dispersed colloidal solution was centrifuged again (3000 rpm, 30 min) and the supernatant was taken to obtain a clear long-term stable dispersed $Co-CsPbBr_3$ colloidal solution.

**Construction of the detection system.** The system consists of two parts: a sample system and a vibration energy level detection system. a schematic diagram of the device is shown in Figure 1b.

Sample system: The Co-CsPbBr$_3$ material and cuvette prepared in this experiment are used as the sample system of the device, and the sample should be higher than 1/3 and lower than 4/5 when adding the sample, so as to ensure that the laser can be irradiated on the sample. When measuring, it is necessary to ensure that the sample system is completely in the magnetic field, so that the material responds to the magnetic field, and the vibration energy level detection system is used to detect.

Vibration energy level detection system: The experimental group uses Raman spectrometer as a reference to make a set of vibration energy level detection device, which is composed of an excitation part, a light processing part, a receiving part and a processing part. Same as the general Raman spectrum, the system uses a laser to excite the sample to produce scattering, and then uses a filter to filter out the elastic scattering, so that the redshift part in the inelastic scattering passes, and the passing light is divided into the light intensity and relative position of the target wavenumber using a grating device, and the system corresponds to the sample system, and the detection wavenumber is 323 cm and 351cm respectively, and there are two detection angles for the central spectral line. The CCD used at the end of the end can send the received optical signal to the computer wirelessly, and the data can be displayed as an image of the light intensity and wavenumber through the python code as detailed in the appendix and the center wavenumber calibration. The magnitude of the measured magnetic field can be calculated by reading the number of target waves and the intensity of the adjacent digits.

**Characterizations.** The structure and morphology of the as-grown samples was characterized using XRD (Bruker D8) and transimission electron microscopy TEM (Talos F200X). UV−vis absorption was measured on a Nicolet Evolution 300 UV/vis spectrophotometer, and PL spectra were taken using a fluorescence spectrometer (AvaSpec-ULS2048CL-EVO). Laser confocal micro-Raman spectrometer (LabRAM HR Evolution) is used to measure the Raman spectrum. The excitation spectra were measured with a fluorescence spectrophotometer (F-4600 FL Spectrophotometer). The Magnetic field generator was tested using a Hall-effect measurement system (Eastchanging ET 9000).

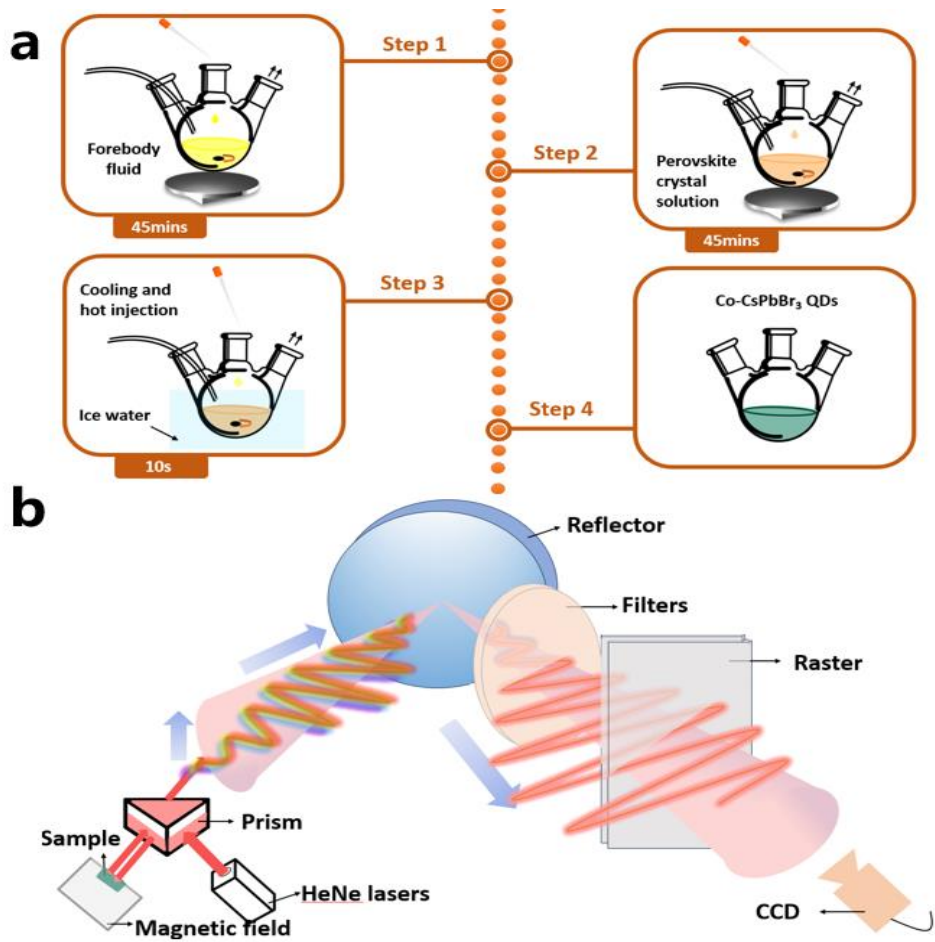

**Figure 1.** a-Diagram of the process of heat injection synthesis of quantum dots; b-Diagram of a weak magnetic field device.

## 3. Results and Discussion

The optimal $CsPbBr_3$ perovskite nanocrystalline structure is that of an octahedron. The alterations in atomic position that occur during doping are illustrated in Figure 2a. Figure 2b illustrates the XRD patterns of the doped $Co$-$CsPbBr_3$ solution and $CsPbBr_3$. The peak position in the image indicates that the $CsPbBr_3$ perovskite exhibits good crystallinity, and the incorporation of Co is evident [15], thereby substantiating the efficacy of the doping process. The crystal structure of $Co$-$CsPbBr_3$ was modelled using software, and as illustrated in Figure 2c, the bonding effect of $Co^{2+}$ and $Br^-$ and the partial substitution of $Pb^{2+}$ by $Co^{2+}$ are evident.

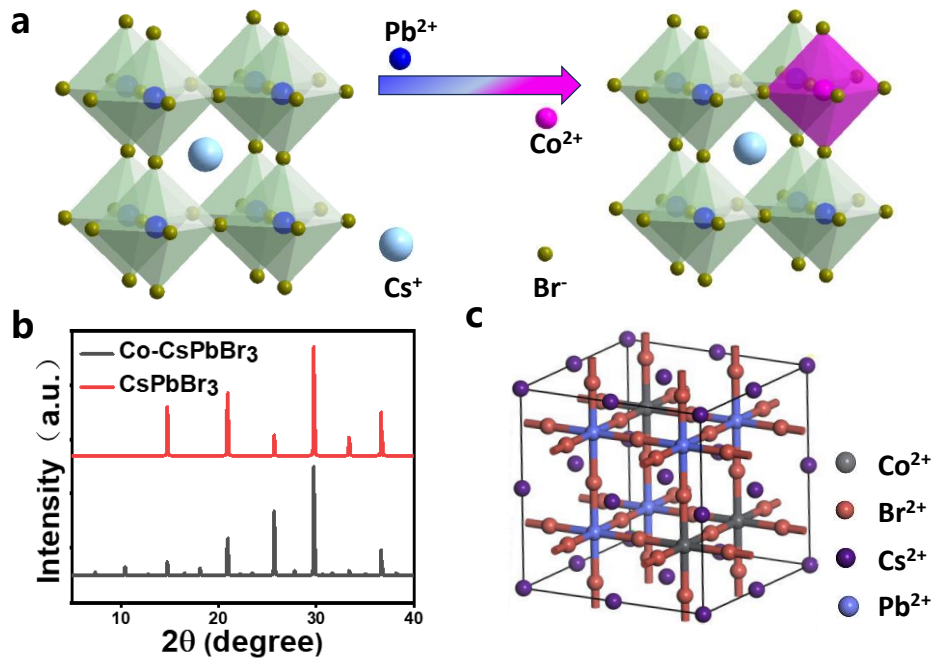

**Figure 2.** a-Simulation process of Co-doped CsPbBr$_3$ Diagram XRD images; b- The Co-CsPbBr$_3$ and CsPbBr$_3$ QDs; c-Simulation model of Co-CsPbBr$_3$.

Figure 3a illustrates the PL spectra of the Co-CsPbBr$_3$ and CsPbBr$_3$ QDs. The CsPbBr$_3$ quantum dots are detected at 521 nm, exhibiting pure green emission, while the Co-CsPbBr$_3$ quantum dot luminescence peak displays cyan luminescence at 500 nm. This evidence substantiates the spectral blue shift of 20 nm that occurs following Co doping. Additionally, the PL quantum yield (PLQY) of Co-CsPbBr$_3$ can be calculated as 71%. According to Figure 3b, the concentration is 6%. Figure 3c illustrates the absorption spectrum of Co-CsPbBr$_3$, which exhibits a primary absorption peak at approximately 490 nm and demonstrates substantial ultraviolet absorption. The Raman spectra of Co-CsPbBr$_3$ are presented in Figure 3d, which contrasts the subsequent changes under the influence of a magnetic field.

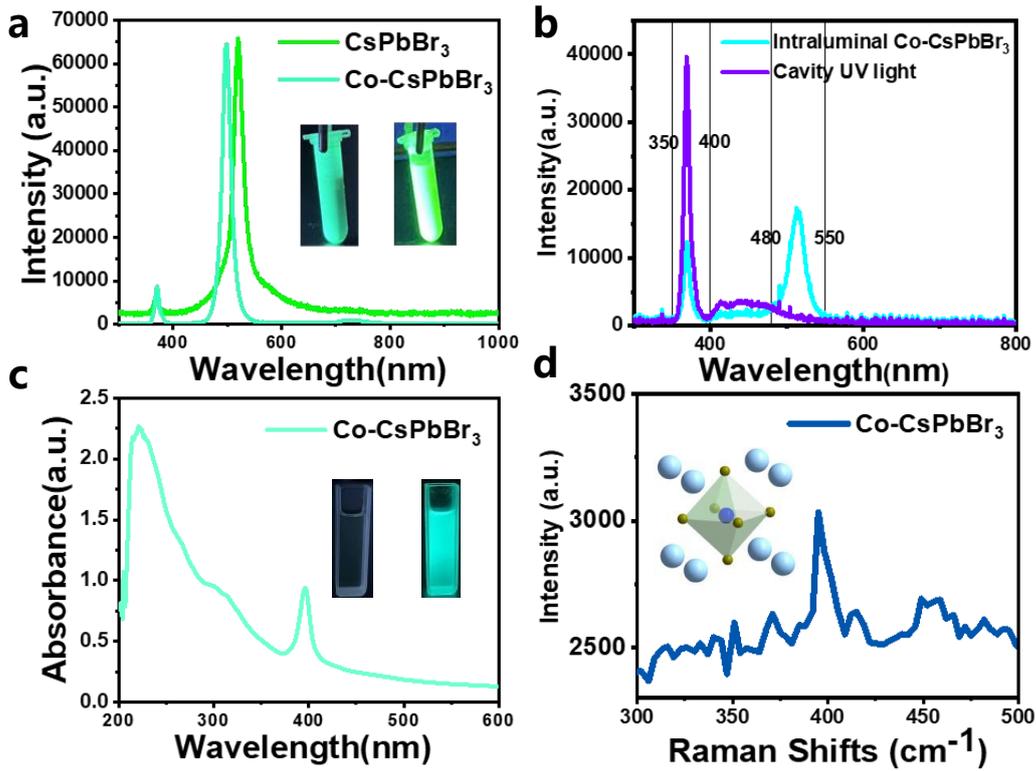

**Figure 3**. a PL spectra of Co-CsPbBr$_3$ and CsPbBr$_3$ b PL spectra of CsPbBr$_3$ in the integrating sphere and cavity c Absorption spectra of Co-CsPbBr$_3$ d Raman spectra of Co-CsPbBr$_3$

Preliminary measurements were made of spectra with a magnetic field of 0 pT and a magnetic field of 950 pT. Due to the attenuation of the laser power, the light intensity of each wavenumber in the spectral pattern decreases, and the power difference between the 0 pt and 950 pT lasers can be calculated in Figure 4a:

$$\Delta P = (I_{950} - I_0) \times A \quad (1)$$

where is A the spot area, I is the light intensity, and $\Delta P$ is the laser power. The error caused by the measured wavenumber minus the attenuation of the laser.

As illustrated in Figure 4a, there is a notable shift in signal intensity at the peak of 323 and 351 cm$^{-1}$, which is close to the bond information [16] of Cs-Br at 318 cm$^{-1}$ and Co-O at 340 cm$^{-1}$. Following extraction, as illustrated in Figure 4b and 4c, it can be observed that the function is essentially linear. Upon fitting the difference, it is postulated that the variation in the intensity of the Raman spectrum reflects the alteration in both the intensity and quantity of the bond vibration. The absence of a change in Co-Br indicates that CoBr may have been separated due to the magnetic field effect of Co bonding with impurity oxygen. This is evidenced by the notable change in the magnetic field conditions, which validates the test method. Conversely, the separation of CoBr also results in a change in the amount of Cs-Br, due to the intensity change of Cs-Br. This indirectly substantiates the efficacy of incorporating chemical changes. The standard data and experimental data of light intensity under different magnetic fields of 323cm$^{-1}$ and 351$^{-1}$ were measured respectively, and the error and deviation were calculated. The data and calculated result curves are shown in figure 4 d-f according to equation (1) and (2).
Error calculation:

$$\delta 1 = \frac{I_{test} - I_{standard}}{I_{standard}} * 100\% \tag{2}$$

Deviation calculation:

$$D = I_1 - ((I_1 + I_2)/2) \tag{3}$$

The error calculations of the test system indicate that the dark box does not provide complete isolation from light, and that there are weak reflections within the system that can influence the measurement results. The sample is susceptible to fluctuations in temperature. When the temperature is elevated, the molecular vibrations of the sample intensify, resulting in a larger measurement outcome. Conversely, when the temperature is reduced, the molecular vibrations of the sample diminish, leading to a smaller measurement result. The optical holder, as designed, exhibits a number of systematic errors resulting from the manufacturing process. Furthermore, following the calibration of the instrument, the positions of the components may be disturbed during use, thereby affecting the accuracy of the data. The replacement of the sample entails a change in the position of the cuvette, which in turn affects the detection position and the reflection of the sample wall. This results in an inaccurate measurement. When the laser is operated for an extended period of time, the power attenuation results in a reduction in the measurement result.

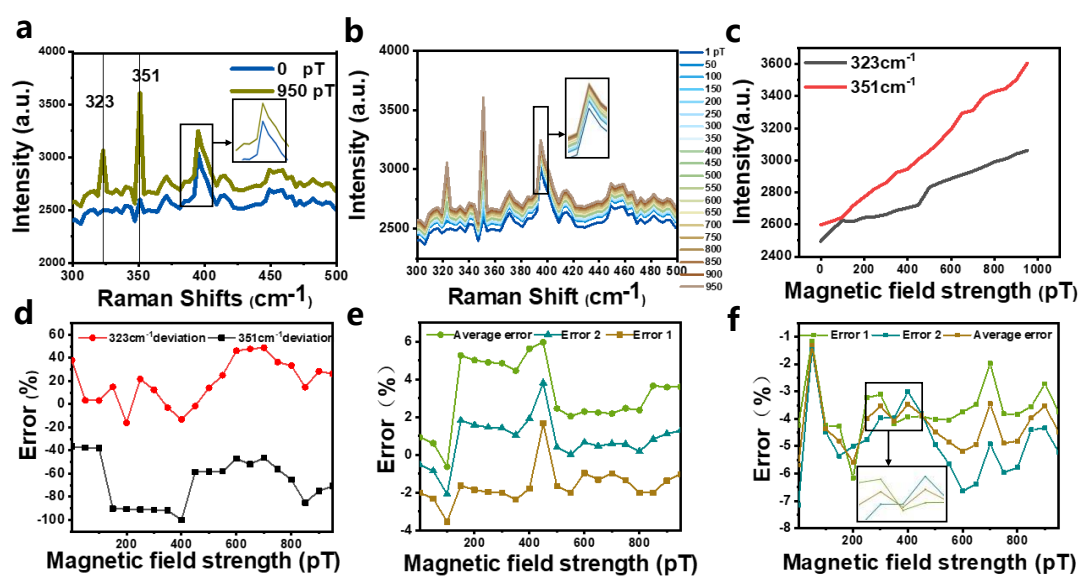

**Figure 4.** a-Raman spectral data at 0 pT and 950 pT magnetic fields; b- Standard data plot of Raman spectra under 0 pT~950 pT magnetic field; c-Magnetic field-light intensity fitting curve; d-Deviation line chart of 323 and 351 cm$^{-1}$; e-323 cm$^{-1}$ error line chart; f-Line chart of error 351cm$^{-1}$.

To facilitate cognitive understanding and study, a simple harmonic approximation was proposed, wherein solely the interaction between the nearest atoms within the molecule was considered. Furthermore, the harmonic oscillator model was adopted as a massless spring linking two balls of the same atomic mass [17]. As illustrated in Figure 5a, the simple harmonic oscillator and the Morse potential energy function exhibit a high degree of similarity in their low-energy vibration dynamics, allowing for the simulation of their respective behaviours. It can therefore be concluded that a harmonic oscillator (resembling a standard spring) is an appropriate means of describing the problem. The left side of Figure 5b depicts the wave function of $\psi_0, \psi_1, \psi_n$, and (the highest), and the right side is the square of the corresponding wave function, which represents the corresponding probability.

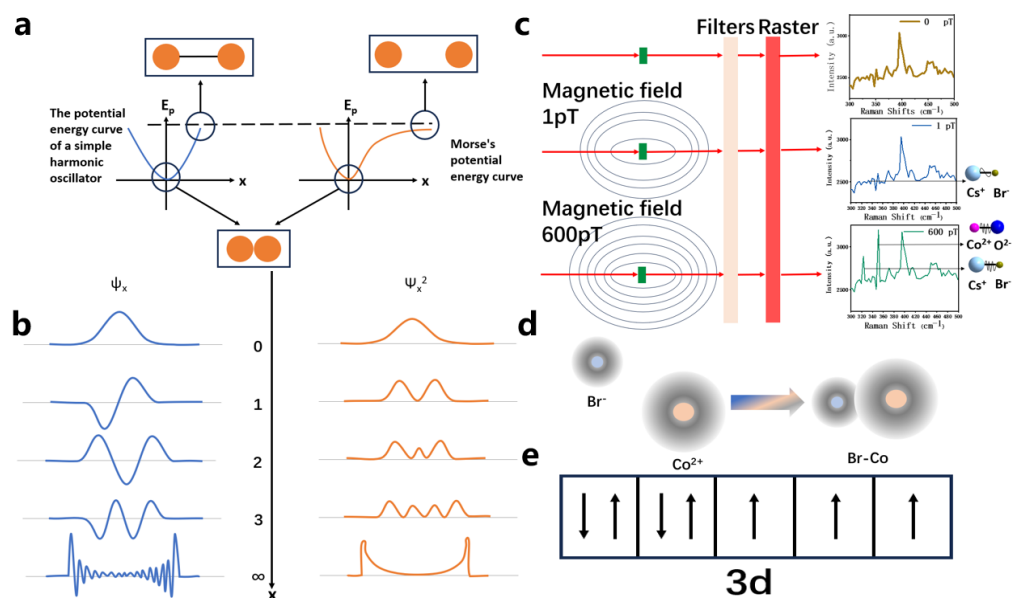

**Figure 5**. a-Schematic diagram of the vibration function and Morse function of the low-level simple harmonic oscillator; b-Schematic diagram of the harmonic oscillator wave function and its square function; c-Introduction to the principle of Raman spectral variation under different magnetic fields; d-The Br-Co Schematic diagram of bonding; e-$Co^{2+}$ diagram of the outermost outer nucleus electron configuration.

The wave function of a chemical bond also has a phenomenon similar to that of the electron wave function: as the energy level increases, so does the node surface of the wave function. In the ground state, the wave function is similar to a normal distribution, with a high middle and decreasing sides. At $\psi_1$, a node appears and the wave function transforms to inversion in the middle [18]. Correspondingly, the square of the wave function of the bond is still high in the middle and low in the ground state, while the $\psi_1$ state appears 0 in the middle and two peaks near the sides. When the chemical bond is in the ground state, the square of the wave function is concentrated in the middle $r_0$ (equilibrium bond length), indicating that the probability of being in equilibrium bond length is the largest in the ground state, and the probability decreases as the bond length increases to both sides [19]. It is conceivable that the two nuclei are relatively stable at the equilibrium bond length, and the farther away from the equilibrium position, the lower the probability. According to the potential energy curve, since the potential energy at the equilibrium position is the lowest, it is the most stable in the ground state (the maximum probability occurs in the equilibrium bond length). When the chemical bond is in the excited state $\psi_1$, the probability of occurrence at the equilibrium bond length is 0, And there are two crests on both sides, and the probability is highest at this time. Since there is a high probability of appearing in a position other than the equilibrium state, the corresponding energy will also be higher than that of the equilibrium state. Therefore, the energy of the $\psi_1$ bond orbital is higher (at the energy corresponding to the position of the second horizontal line). It should also be noted that outside the scope of the second line $[r_1, r_1^{'}]$, is still a chance that the key will appear. Therefore, at state $\psi_1$ the keys don't just vibrate within the range $[r_1, r_1^{'}]$, rather, it's a much wider range. It's just that at this time, the energy of the bond level is just corresponding to the energy of $[r_1, r_1^{'}]$. It should also be noted that in the simple harmonic oscillator, the average bond length of the $\psi_1$ and $\psi_0$ bonds is $r_0$. However, since the potential energy of the chemical bond is a function of Morse potential energy, the curve is skewed to the right, so the equilibrium bond length is actually larger than $\psi_0$. That is, when the bond is excited, the average bond length of the chemical bond increases. When the vibrational energy level of the bond is getting higher and higher, after reaching $\psi_n$, there should be n nodes at this time, and the wave function is shown on the left side of Figure 5b. The middle amplitude is low, while the outermost amplitude is the largest. The square of the wave function is shown on the right side of Figure 5b, where the bond is likely to be very long or very short, and the closer it is to the equilibrium bond, the lower the probability of bond length. Since the bond is likely to appear in a very long or short state, the energy level is very high; And the higher the energy level of the bond, the farther the peak of the maximum probability of the bond is. This is close to the spring oscillator, which has a velocity of 0, maximum acceleration, highest energy and longest residence time at the longest and shortest.

Magnetic fields cause different bonds to change, and bond changes lead to spectral changes, as shown in figure 5c. No change in Co-Br was found here, which indicates that CoBr may have been separated, on the one hand, due to the magnetic field effect of Co bonding with impurity oxygen, so the change is very obvious under different magnetic field conditions, which proves the validity of the test method; On the other hand, since CoBr is separated, the number of Cs-Br changes, because Cs-Br has an intensity change, which indirectly proves the effectiveness of chemical change incorporation. Based on the understanding of bonds in quantum mechanics and the understanding of bonds by traditional spring oscillators, the difference is that the energy level of the bond generated by quantization [20] and the corresponding bond length probability determine the state of the bond, and the bond length is probabilistic at this time. In the ground state, the bond length is close to a fixed (or a slight wobble), and at the highest energy level, it is closest to the state in which a spring oscillator appears, which is called vibration. Since there is no spring between the two atoms, there is no such thing as "vibration", which is essentially caused by the density of the bond length.

Chemical bond is the result of atomic interaction, its essence is the redistribution of electrons around the atom in space after bonding, that is, the deformation of the electron orbital, this distribution makes the energy in the system decrease, is an important part of maintaining the existence of compounds, the $CoBr_2$ used in this experiment, the bond is essentially an ionic bond, but because cobalt is a transition metal, so the bond with bromine is an incomplete ionic bond, which has a part of the covalent bond nature. Ionic bonds are formed due to the difference in the positive and negative charges carried by two atoms, and the electrostatic forces between the atoms are attracted to each other. Covalent bond is the overlap of two atomic orbitals to form two molecular orbitals: bonding orbitals and antibonding orbitals, when the molecule is in the ground state, the two electrons mainly move in the bonding orbitals, forming a negative ion cluster to bind and connect two positively charged atoms. The incomplete ionic bond in $CoBr_2$ is that some atomic orbitals overlap when the +2 valent cobalt ion and the negative monovalent bromide ion interact with each other, thus having some of the same properties as the covalent bond [21]. A schematic diagram of Br-Co bonding is shown in Figure 5d.

In chemistry, the average distance between two atoms in the fundamental energy state is defined as the bond length, and the angle between two chemical bonds connected to the same atom in the fundamental energy state is defined as the bond angle, that is, the average angle between two atoms connected to the same atom in the stable state [22]. A chemical bond is not a connection of an entity, but a binding force of a negative charge group generated by the movement of electrons in a bonding orbital to two atoms. The atoms themselves are still in constant motion, and the distance between the two atoms is different in every instant, but their average distance is a fixed value. Therefore, the bond length in the unexcited state is the equilibrium bond length, because the energy absorbed by the molecule has the characteristics of quantization, so the molecule can only absorb the energy of the difference between the two energy levels, so the energy change of the molecule is also discontinuous, so the energy of the molecule is also quantized step, that is, the molecular energy level (valence electron energy level). When the valence electron energy level of a molecule changes, its vibrational energy level and rotational energy level must change together[15], and the phenomenon of this experimental sample is the change of vibrational energy level.

The reason why $Co^{2+}$ is ferromagnetic at room temperature involves the properties of its electronic structure and magnetic moment [23]. The ionic structure of $Co^{2+}$ is 25 electrons and is arranged in $1s^22s^22p^63s^23p^63d^7$. This arrangement results in the formation of unpaired electrons around the electronic structure of $Co^{2+}$ around its nucleus which can generate a magnetic moment which makes $Co^{2+}$ magnetic [24]. The outermost outer nuclear electron configuration diagram is shown in Figure 5e.

In $Co^{2+}$, the electrons of the 3d orbital play a key role in magnetism. Due to the geometry of the 3d orbitals, the unpaired electrons in $Co^{2+}$ can produce magnetic moments, which makes $Co^{2+}$ magnetic. In addition, the interaction between $Co^{2+}$ also strengthens its magnetic properties. At room temperature, the electronic structure of $Co^{2+}$ keeps it in a ferromagnetic state, the magnetic moments tend to align in the same direction [25]. This arrangement results in $Co^{2+}$ exhibiting magnetism under the action of an applied magnetic field and remaining in this state at room temperature. When $CoBr_2$ molecule forms complexes or complexes with other molecules, the Co-Br bond changes so that it can exhibit ferromagnetism. S-S'-1,2-sulfide benzene reacts with $CoBr_2$ to form a cobalt complex [(bptp)$CoBr_2$], which has been shown to act as a single-molecule magnet and exhibit ferromagnetism. According to the magnetic behavior of

[(bptb)CoBr$_2$] in the temperature range of 2 to 300 K [26], at 300 K, the χ MT value is 2.91 cm3 Kmol$^{-1}$, which is much higher than the spin value of 1.875 cm3 Kmol$^{-1}$ of single-core high-spin Co$^{2+}$ ions (S=3/2, g=2.0). This value is in the range of anisotropic high spin Co$^{2+}$ centers of 2.1 to 3.4 cm3Kmol$^{-1}$ that have been reported attributed to orbital contributions [27]. In addition, the χ MT value gradually decreases from 300 K to 130 K, and then rapidly decreases below 130 K, reaching a minimum of 1.82 cm$^3$ Kmol$^{-1}$ at 2 K. The decrease in χ MT values at low temperatures can be attributed to the intrinsic magnetic anisotropy of Co$^{2+}$ ions. In this experiment, CoBr$_2$ was incorporated into CsPbBr$_3$ using the similarity principle, in which CoBr$_2$ formed a complex with other molecules, and the Co-Br bond was changed to exhibit ferromagnetism.

## 4. Conclusion

In this study, we employed the method of hot injection to synthesis Co-CsPbBr$_3$ quantum dots and utilize the data pertaining to their structural alterations subsequent to exposure to a magnetic field for the purpose of conducting weak magnetic field measurements. The alterations in atomic position that occur during doping Co-CsPbBr$_3$ model has been presented. The XRD patterns of the doped Co-CsPbBr$_3$ QDs indicates a good crystallinity, and the successful incorporation of Co is evident. The Co-CsPbBr$_3$ QDs PL peak displays cyan luminescence at 500 nm. This evidence substantiates the spectral blue shift of 20 nm that occurs following Co doping. Additionally, the PL quantum yield of Co-CsPbBr$_3$ was calculated as 71%. The notable shift in signal intensity at the peak of 323 and 351 cm$^{-1}$ confirmed by Raman spectrum. The magnetic field strength was calibrated against the Raman signal, showing a linear relationship at the pT level. In summary, this paper presents a novel approach to weak magnetic field measurement, utilising the Co-CsPbBr$_3$ material to acquire vibrational energy level information. In comparison with conventional weak magnetic field measurement techniques, this approach enables the accurate detection of weak magnetic fields with a sensitivity of 1 pT ($10^{-8}$ gauss). Furthermore, the compactness of the measurement apparatus can be enhanced by utilising the zero-dimensional perovskite material.

## Acknowledgments

National Natural Science Foundation of China (62305262). Natural Science Foundation of Shaanxi Province (2022JQ-652). Shaanxi Fundamental Science Research Project for Mathematics and Physics (22JSQ026).

## Data availability

The data supporting the findings of this study are available in the paper, Supplementary Information, as well as from the corresponding authors upon request.

## Competing interests

The author declares no competing interests.